\documentclass[iop,preprint]{emulateapj}
\usepackage{graphicx,times}

\slugcomment{The Astrophysical Journal Letters, 781:L40, 2014 January 15}

\shorttitle{HST observations of ram-pressure stripping in MACS clusters}
\shortauthors{Ebeling, Stephenson, \& Edge}

\begin{document}

\title{JELLYFISH: EVIDENCE OF EXTREME RAM-PRESSURE STRIPPING IN MASSIVE GALAXY CLUSTERS\footnote{Based on observations made with the NASA/ESA \textit{Hubble Space Telescope}, obtained at the Space Telescope Science Institute, which is operated by the Association of Universities for Research in Astronomy, Inc., under NASA contract NAS 5-26555. These observations are associated with programs GO-10491, -10875, -12166, and -12884.}
}

\author{H.\ Ebeling, L.N.\ Stephenson}
\affil{Institute for Astronomy, University of Hawaii, 2680 Woodlawn Drive, Honolulu, HI 96822, USA}
\and
\author{A.C.\ Edge}
\affil{Institute for Computational Cosmology, Department of Physics, University of Durham, South Road, Durham, DH1 3LE, UK}

\begin{abstract}
Ram-pressure stripping  by the gaseous intra-cluster medium has been proposed as the dominant physical mechanism driving the rapid evolution of galaxies in dense environments. Detailed studies of this process have, however, largely been limited to relatively modest examples affecting only the outermost gas layers of galaxies in nearby and/or low-mass galaxy clusters. We here present results from our search for extreme cases of gas-galaxy interactions in much more massive, X-ray selected clusters at $z>0.3$. Using \textit{Hubble Space Telescope}  snapshots in the F606W and F814W passbands, we have discovered dramatic evidence of ram-pressure stripping in which copious amounts of gas are first shock compressed and then removed from galaxies falling into the cluster. Vigorous starbursts triggered by this process across the galaxy-gas interface and in the debris trail cause these galaxies to temporarily become some of the brightest cluster members in the F606W passband, capable  of  outshining even the Brightest Cluster Galaxy.  Based on the spatial distribution and orientation of systems viewed nearly edge-on in our survey, we speculate that infall at large impact parameter gives rise to particularly long-lasting stripping events. Our sample of six spectacular examples identified in clusters from the Massive Cluster Survey, all featuring $M_{\rm F606W}<-$21 mag, doubles  the number of such systems presently known at $z>0.2$ and facilitates detailed quantitative studies of the most violent galaxy evolution in clusters. 
\end{abstract}


\section{Introduction}
That galaxies evolve in both color (from blue to red) and morphology (from late to early Hubble types) is a central paradigm of galaxy formation and hierarchical evolution that is backed by abundant observational evidence \citep[e.g.,][]{2004ApJ...608..752B,2007ApJ...665..265F}. Just how this evolution comes to pass is, however, still a subject of intense debate and investigation. Early work by \citet{1980ApJS...42..565D} established the importance of environment for the morphological evolution of galaxies: since early-type galaxies are essentially absent in the field but dominant in the cores of rich clusters, the group and cluster environment must be instrumental in the transformation of spirals into lenticular and elliptical galaxies.

Several physical processes have been proposed as drivers of this transformation. Galaxy-galaxy mergers \citep{1972ApJ...178..623T} have been found to dominate the evolution of galaxies in low-density environments \citep[e.g.][]{2000MNRAS.311..565L} where relatively low relative velocities result in a high cross section for mergers. In high-density environments several mechanisms compete, the primary ones being ram-pressure stripping \citep{1972ApJ...176....1G,1983ApJ...270....7D}, galaxy ``harassment" \citep{1996Natur.379..613M,1998ApJ...495..139M}, and tidal compression \citep{1990ApJ...350...89B}. 

In massive clusters, ram-pressure stripping is expected to be  by far the most efficient of these processes.  For a given galaxy, ram pressure is directly proportional to the density of the intracluster medium (ICM) and to the square of the galaxy's velocity relative to the ICM.  Extensive numerical simulations  predict that gradual stripping should be pervasive even in low-mass clusters \citep[e.g.][]{2001ApJ...561..708V}; in the most massive clusters, the environment encountered by infalling galaxies can lead to complete stripping of their gas content in a single pass through the cluster core \citep[e.g.][]{1984MNRAS.208..261T,1999MNRAS.308..947A,2009A&A...499...87K,2012A&A...544A..54S}. 

\begin{figure*}[t]
\includegraphics[width=0.33\textwidth]{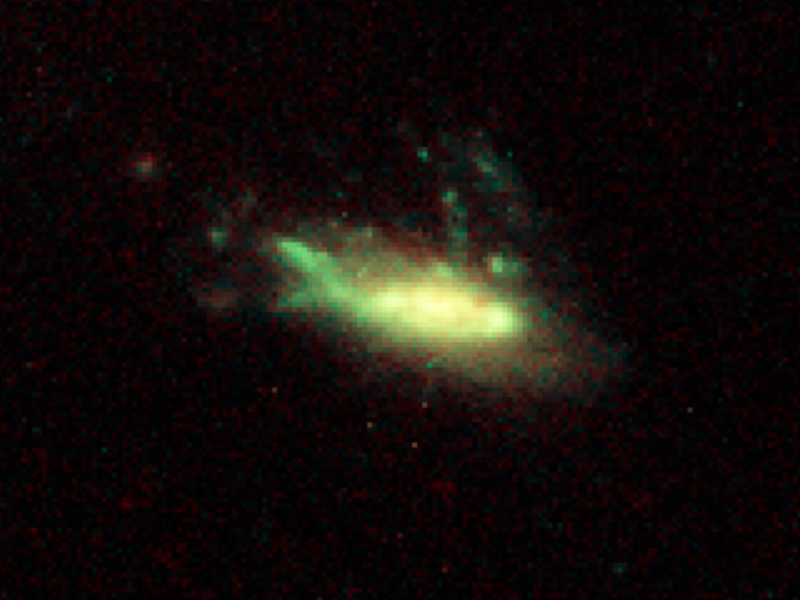}
\includegraphics[width=0.33\textwidth]{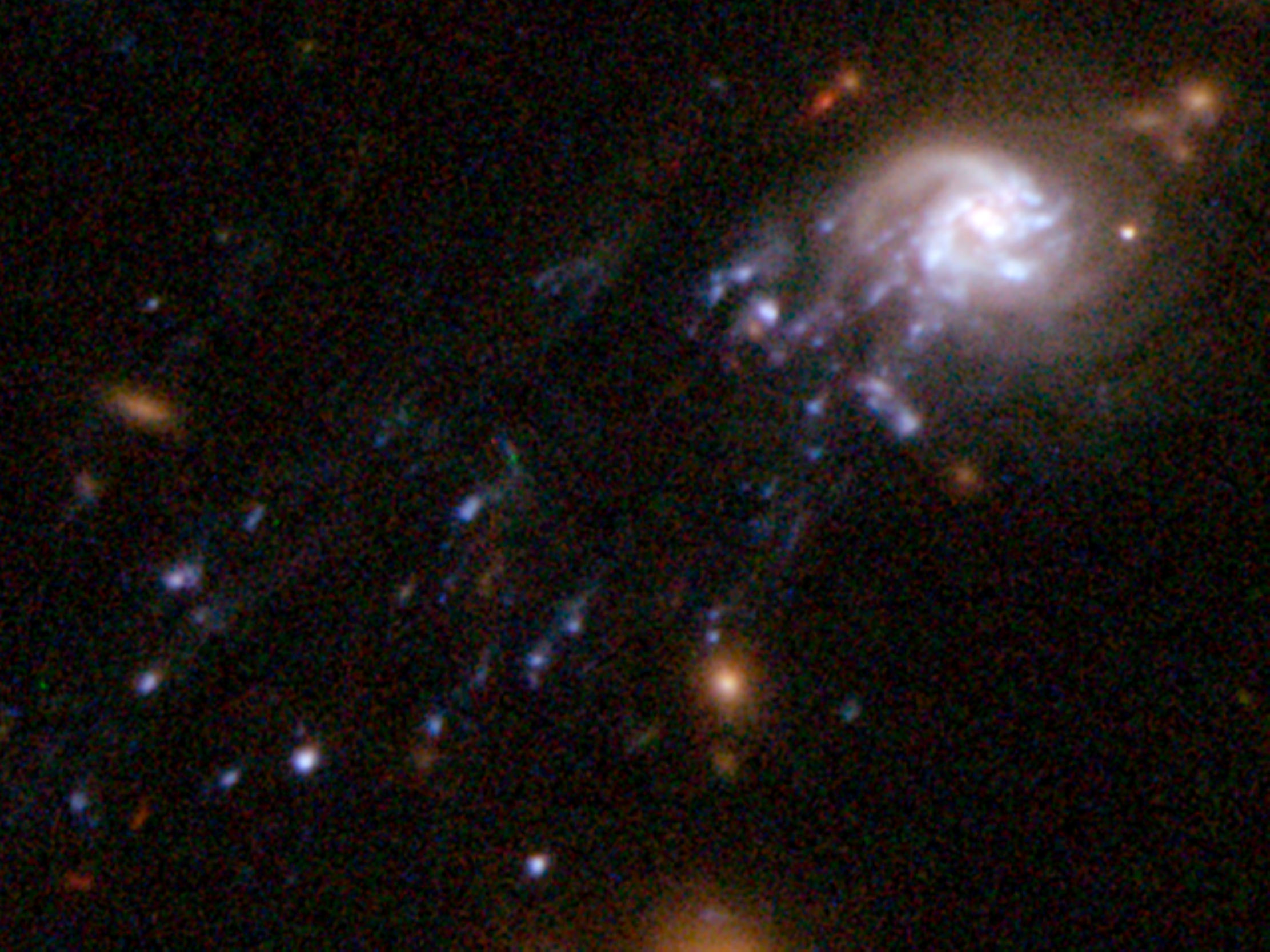}
\includegraphics[width=0.33\textwidth]{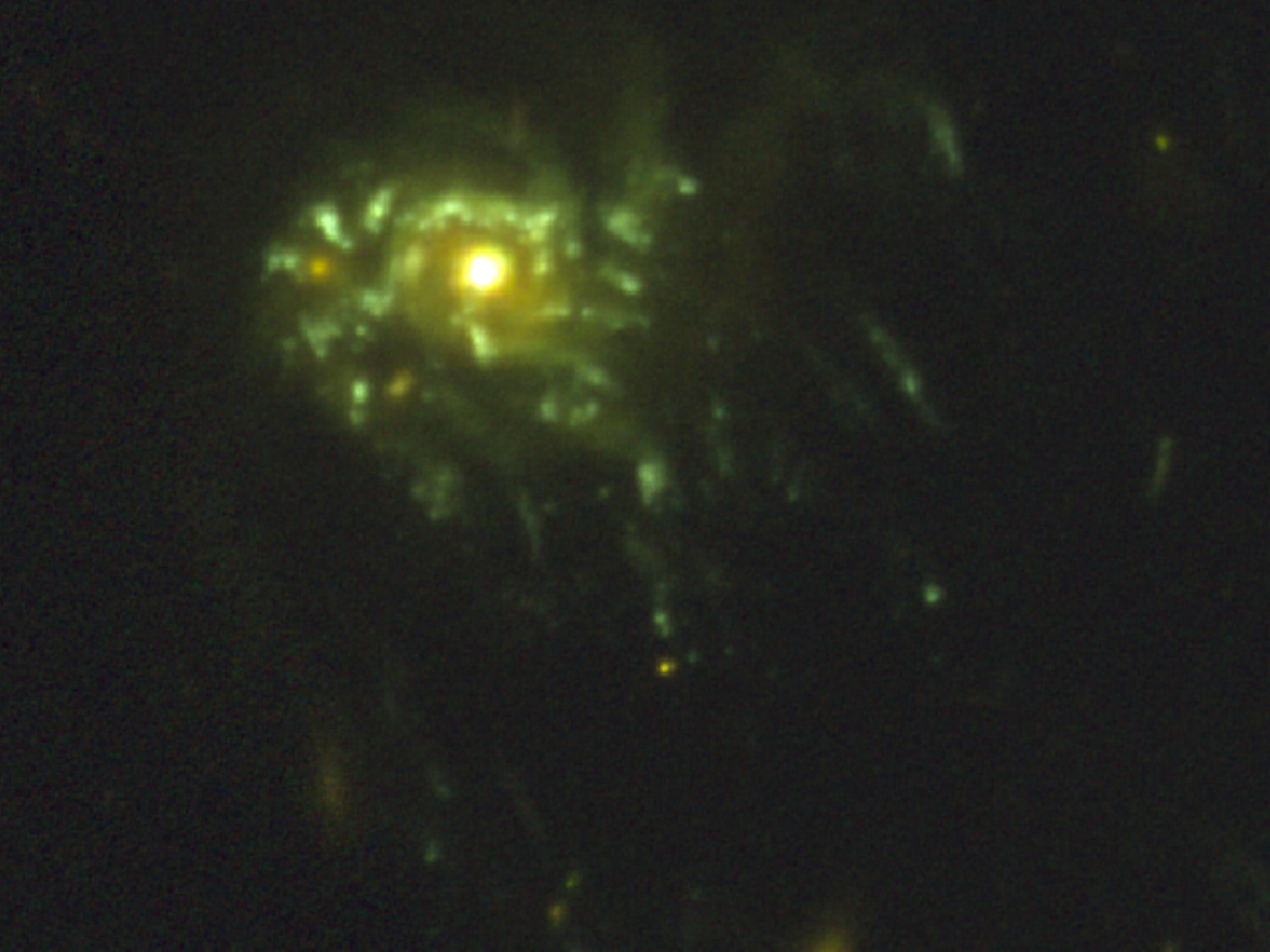}
\caption{\textit{HST} images of extreme cases of ram-pressure stripping in galaxy clusters at $z>0.2$. From left to right: galaxy C153 in A2125 at $z=0.20$ \citep[WFPC2, F606W+F814W,][]{2006AJ....131.1974O}; galaxy 234144--260358 in A2667 at $z=0.23$ \citep[ACS, F450W+F606W+F814W,][]{2007MNRAS.376..157C}; galaxy F0083 in A2744 at $z=0.31$ \citep[ACS, F435W+F606W+F814W,][]{2012ApJ...750L..23O}.}\label{fig:jelly_lit}
\end{figure*}

\section{The observational signature of ram-pressure stripping}

Since detailed investigations of galaxy transformations require high spatial resolution, observational studies have so far focused on galaxies in nearby clusters, most prominently Virgo, Coma, and A1367 (all at $z<0.03$). Imaging and spectral data collected over a wide range of wavelengths yielded a wealth of information on the observational signature of galaxy transformations in clusters. Specifically, a considerable fraction of spiral galaxies in clusters were found to be significantly deficient in H{\sc i} and to exhibit asymmetric morphologies. In addition, they show enhanced star formation in compressed regions, but reduced or fully quenched star formation in the outer disks \citep[][]{1985ApJ...292..404G,1989ApJ...346...59G,1990AJ....100..604C,1994AJ....107.1003C,2006ApJ...651..811B}. Finally, a gaseous tail with embedded bright knots of star formation was observed in the wake of some infalling galaxies \citep{2008ApJ...688..918Y,2007ApJ...671..190S,2010ApJ...716L..14H,2010AJ....140.1814Y}. All of these features are expected in a scenario in which galaxies being accreted from the field experience ram-pressure stripping upon entering the cluster environment and cannot easily be explained by competing physical mechanisms.

Although it has been questioned whether ram-pressure stripping can ultimately transform spiral into lenticular galaxies  \citep[see, e.g.,][]{2006PASP..118..517B}, it is thus clear that galaxy--gas interactions play a central role in the evolution of galaxies in dense environments. Past studies of the relevant physical processes were, however, mainly and necessarily based on observations of modest stripping events. Extreme ram-pressure stripping is expected to proceed rapidly and likely requires both high ICM densities and suitable galaxy properties (e.g., favorable infall trajectory, gas mass, orientation), conditions that are unlikely to be met in the small number of nearby clusters, all of which feature relatively low mass (except for Coma). Indeed, observations show atomic hydrogen in infalling galaxies to be displaced and partly removed \citep[e.g.,][]{2010MNRAS.403.1175S}, but find the denser, more centrally located molecular gas essentially unperturbed \citep[e.g.,][]{1997A&A...327..522B,2001A&A...374..824V}. In addition, star formation is found to be globally quenched (and only mildly enhanced in compressed regions) rather than massively boosted at the galaxy--ICM interface. All of these findings point toward mild ram-pressure stripping acting gradually and repeatedly on galaxies falling into, or orbiting in, clusters. 

Extremely rapid and essentially complete stripping must occur too, but is much more rarely observed because of the, presumably, much shorter duration of the event ({\bf$\mathbf{{<}10^8}$ yr}, i.e., a fraction of a crossing time) and because of its reliance on a truly extreme environment. The latter is, however, routinely encountered by galaxies falling into very massive clusters where the particle density\footnote{In spite of the enrichment of the ICM with metals, hydrogen is the dominant atomic species encountered; hence, an ICM particle density of $10^{-3}$ cm$^{-3}$ corresponds approximately to a mass density of 10$^{-24}$ g cm$^{-3}$.} of the ICM  easily exceeds 10$^{-3}$ cm$^{-3}$, and peculiar galaxy velocities of 1000 km s$^{-1}$ or more are common. 

Consistent with the aforementioned observational bias, the most dramatic examples of ram-pressure stripping discovered so far were found in moderately distant, X-ray luminous clusters. Shown in Figure~\ref{fig:jelly_lit} are the three most dramatic cases of ram-pressure stripping discovered so far in \textit{Hubble Space Telescope (HST)} images of clusters at $z>0.2$  \citep{2006AJ....131.1974O,2007MNRAS.376..157C,2012ApJ...750L..23O}. In all cases, the respective galaxy features intense star formation across much of its visible disk, making it the brightest  member of its host cluster at 4000\AA. Although debris trails of star-forming knots are discernible already with WFPC2 (left panel of Figure~\ref{fig:jelly_lit}), the greatly superior resolution and sensitivity of the Advanced Camera for Surveys (ACS) is evident (central and right panel of Figure~\ref{fig:jelly_lit}). Detailed studies of all three objects suggest that multiple phases of ram-pressure stripping can overlap sufficiently to be observed concurrently: shock compression of the ISM at the galaxy-gas interface causing vigorous and widespread starbursts, removal of intragalactic gas,  star formation in molecular clouds swept out of the galaxy, as well as partial back-infall. Furthermore, tidal compression in the cluster's gravitational potential may contribute to the observed pronounced and wide-spread star formation.

More robust conclusions are hard to arrive at from the few examples observed to date since, as expected from simple theoretical considerations and the results of numerical simulations, the progression and observational signature of extreme ram-pressure stripping depends greatly on the intrinsic properties, orientation, and orbital parameters of the infalling galaxy. A significantly larger sample of galaxies caught in this violent phase of their evolution is needed to allow us to test, on a sound statistical basis, the predictions of numerical simulations in a physical regime that has barely been probed in studies of galaxy evolution in nearby clusters.
 
\section{Sample, observations, and data reduction}

In order to identify additional examples of extreme galaxy-gas interactions in very massive clusters we searched for the tell-tale signature of ram-pressure stripping in images of clusters from the Massive Cluster Survey \citep[MACS;][]{2001ApJ...553..668E} obtained with ACS aboard \textit{HST}.  Our project uses all 37 MACS clusters\footnote{Four of these in fact hail from the southern extension of MACS which covers the extragalactic sky at $\delta<-40^\circ$.} observed with ACS in two passbands (F606W and F814W) as part of  the \textit{HST} snapshot programs GO--10491, --10875, --12166, and --12884 (PI: Ebeling) as of 2013 June 1.  This sample constitutes an unbiased subset of the larger SNAP target list of 128 MACS clusters at $\mathbf{0.3{<}z{<}0.5}$, since their selection for observation was solely driven by constraints on the \textit{HST} observing schedule. Charge-transfer inefficiency corrected images were aligned and registered using the astrometric solution of the F606W image as a reference; we created false-color images using the average of both bands for the green channel. Source properties were determined using SExtractor \citep{1996A&AS..117..393B} in dual-image mode with F606W chosen as the detection band.  

Our search for galaxies experiencing violent encounters with the ICM consists of two parts. We first perform a simple visual inspection of the color images of all clusters to identify the brightest and most spectacular examples of extreme ram-pressure stripping. The second phase then uses the unambiguous cases thus unveiled as a training set to establish quantitative color and morphology criteria that allow the selection of fainter objects of conspicuous but less compelling visual appearance to create an even larger sample of galaxies that might be experiencing a similar transformation. In this Letter, we focus on the former step; a detailed description of the second phase and the resulting candidate list will be presented in a separate paper (in preparation).

Since the clusters in our sample cover a range of redshifts, $0.3<z<0.5$, the metric scale of our images varies between images from 4.45 to 6.10 kpc arcsec$^{-1}$ and the field of view of ACS just covers an inscribed circle of radius 450 to 617 kpc. We assume the concordance $\Lambda$CDM cosmology ($\Omega_{\rm M}=0.3$, $\Omega_{\rm \Lambda}=0.7$) and H$_0$=70 km s$^{-1}$ Mpc$^{-1}$.

\begin{figure*}[t]
\includegraphics[width=0.5\textwidth,page=1,clip,trim=15mm 5mm 25mm 15mm]{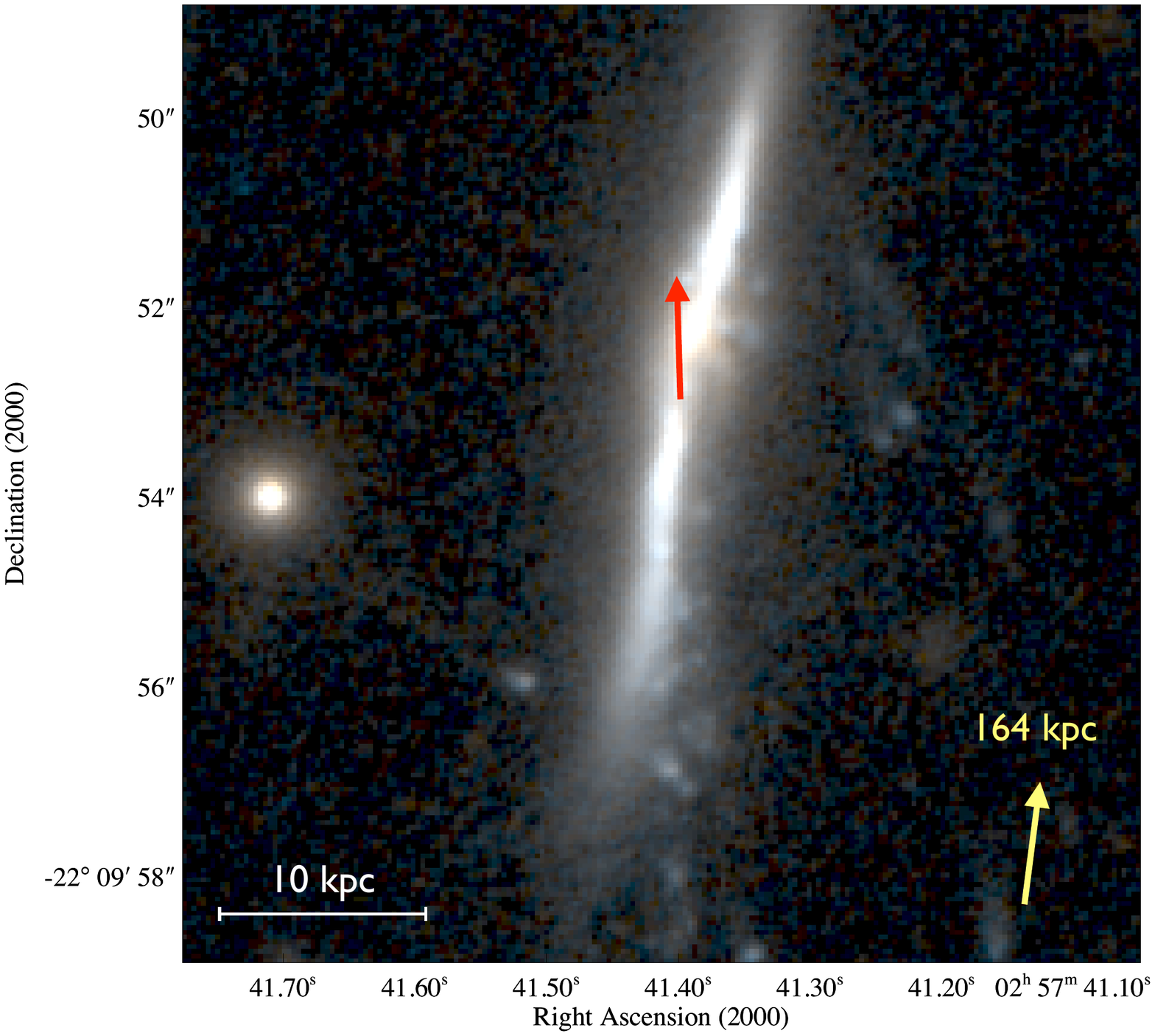}
\includegraphics[width=0.5\textwidth,page=2,clip,trim=15mm 5mm 25mm 15mm]{im_arrows}\\[-1mm]
\includegraphics[width=0.5\textwidth,page=3,clip,trim=15mm 5mm 25mm 15mm]{im_arrows}
\includegraphics[width=0.5\textwidth,page=4,clip,trim=15mm 5mm 25mm 15mm]{im_arrows}\\[-1mm]
\includegraphics[width=0.5\textwidth,page=5,clip,trim=15mm 5mm 25mm 15mm]{im_arrows}
\includegraphics[width=0.5\textwidth,page=6,clip,trim=15mm 5mm 25mm 15mm]{im_arrows}\\[-5mm]
\caption{\textit{HST} images (F606W+F814W) of extreme cases of ram-pressure stripping in MACS galaxy clusters at $0.30<z<0.43$. In each panel, the direction and projected distance to the cluster center (as given by the location of the BCG) is marked in the bottom, right corner; red arrows denote the approximate direction of motion of the respective galaxy.}\label{fig:jelly}
\end{figure*}

\section{Jellyfish galaxies: extreme ram-pressure stripping}

Our visual inspection of the ACS color images of all 37 MACS clusters in our sample uses three primary criteria to identify galaxies undergoing extreme ram-pressure stripping: (1) a  strongly disturbed morphology indicative of unilateral external forces; (2) a pronounced brightness and color gradient suggesting extensive triggered star formation; and (3) compelling evidence of a debris trail. In addition, the directions of motion implied by each of these three criteria have to be consistent with each other. The resulting sample, clearly biased in favor of galaxies moving in, or close to, the plane of the sky, constitutes the set of cases we consider unambiguous; the larger set of galaxies passing at least two of the above criteria is retained for training purposes for the second phase of our project.

We show in Figure~\ref{fig:jelly} the brightest galaxies classified by us as systems experiencing extreme, text-book ram-pressure stripping; key properties of these objects are listed in Table~\ref{tab:jelly}. For obvious reasons we shall, in the following, refer to them as ``jellyfish" galaxies. 

\subsection{Morphology and Brightness}

Although all of our jellyfish galaxies share by design the morphological characteristics by which they were selected, differences between them are clear from Figure~\ref{fig:jelly}. The extent and strength of star formation varies, as does the degree of morphological deformation. The latter is likely primarily caused by differences in the inclination of the disk to the direction of motion, whereas the former may be indicative of the phase of the transformation, in the sense that star formation at the gas-galaxy interface can be expected to fade as gas is removed. 

For at least one of our galaxies, MACSJ1258-JFG1, a significant fraction of the observed flux originates from nuclear emission which was likely triggered or at least boosted by ram-pressure induced influx of gas onto the nucleus. Indeed, MACSJ1258-JFG1 is known to host an active galactic nucleus (AGN) and is classified as a QSO in the literature (SDSS J125759.49+470245). AGN emission was also reported by \citet{2012ApJ...750L..23O} for galaxy F0083 in A2744, shown in the rightmost panel of Figure~\ref{fig:jelly_lit}.
 
The complex and variable morphology of jellyfish galaxies frequently causes them to be classified as blends or superpositions of several objects in SExtractor's source list. We  compute total magnitudes by adding the isophotal flux from all components of the source as identified from the SExtractor segmentation map; results are listed in Table~\ref{tab:jelly}. At absolute magnitudes often exceeding $M_{\rm F606W}=-$22 (also quoted in Table~\ref{tab:jelly}) these galaxies are among the brightest members of their respective host clusters. In fact, MACSJ0451-JFG1 temporarily even outshines the Brightest Cluster Galaxy (BCG) by 0.4 mag in the near-UV and blue part of the electromagnetic spectrum.

\subsection{Spatial Distribution}

Although the position of our jellyfish candidates is only known in projection, we can compare their location relative to the cluster center with that of galaxies on the cluster red sequence. We find the two distributions to be statistically indistinguishable. While this result alone does not allow us to put meaningful constraints on the duration of extreme stripping events, we note that the observed small projected distances (90--360 kpc; Figure~\ref{fig:jelly}) of our jellyfish galaxies from the center of their host cluster\footnote{These small distances from the cluster center cannot be purely a selection effect since, for clusters at $z{=}$0.3--0.5, the field of view of \textit{HST}/ACS extends to typically 0.5 Mpc (radius).}  are hard to reconcile with the simple picture in which stripping proceeds rapidly once an infalling galaxy passes inside the ram-pressure stripping radius\footnote{Defined as the radius at which stripping becomes efficient in a galaxy resembling the Milky Way (Ma et al.\ 2008).} ($\sim$1 Mpc).

Additional clues about the three-dimensional trajectories of these systems can be gleaned from the distribution of the directions of motions implied by the orientations of the debris trails (red arrows in Figure~\ref{fig:jelly}). The deduced projected velocity vectors do, in general not coincide with the direction toward the cluster center (also shown in Figure~\ref{fig:jelly}), in contrast to the findings of \citet{2010MNRAS.408.1417S} who found galaxies experiencing ram-pressure stripping in the Coma Cluster to occupy primarily radial orbits. Although any conclusions have to remain tentative given the still small size of our sample and the, compared to the study of  \citet{2010MNRAS.408.1417S},   limited radial range probed by our imaging data, tangential trajectories with large impact parameters appear common, most likely as a result of our explicit focus on systems moving close to the plane of the sky.\footnote{Interestingly, galaxy 234144--260358 in A2667 also appears to pass the cluster center at large impact parameter \citep[][see also Figure~\ref{fig:jelly_lit}]{2007MNRAS.376..157C}.}  Our project may thus have unveiled a population of galaxies whose infall trajectories give rise to particularly dramatic and long-lived stripping events.

\begin{deluxetable}{l@{\hspace{2mm}}c@{\hspace{2mm}}c@{\hspace{2mm}}c@{\hspace{2mm}}c}
\tabletypesize{\scriptsize}
\tablecaption{Positions, Host Cluster Redshifts, Apparent F606W Magnitudes, and Absolute F606W Magnitudes for the Galaxies Shown in Figure~\ref{fig:jelly}. \label{tab:jelly}}
%\tablewidth{0pt}
\tablehead{\colhead{Name} & \colhead{R.A.\ (J2000) Decl.}  & \colhead{$z_{\rm cl}$} & \colhead{$m_{\rm gal}$} & \colhead{$M_{\rm gal}$}}
\startdata
MACSJ0257-JFG1 & 02 57 41.4 \,\,$-$22 09 53 & 0.320 & 18.4 & $-$22.7\\ 
MACSJ0451-JFG1 & 04 51 57.3 \,\,$+$00 06 53 & 0.429 & 19.6 &$-$22.3\\
MACSJ0712-JFG1 & 07 12 18.9 \,\,$+$59 32 06 & 0.328 & 19.0 & $-$22.2\\ 
MACSJ0947-JFG1 & 09 47 23.1 \,\,$+$76 22 52 & 0.354 & 19.8 & $-$21.6\\ 
MACSJ1258-JFG1 & 12 57 59.6 \,\,$+$47 02 46 & 0.331 & 18.6 & $-$22.6\\
MACSJ1752-JFG1 & 17 51 56.1 \,\,$+$44 40 20 & 0.364 & 20.2 & $-$21.3
\enddata
\end{deluxetable}

\section{Summary}

Extreme cases of ram-pressure stripping in which the majority of the intra-galactic gas is rapidly removed from a galaxy falling into a massive cluster are predicted by theoretical considerations and expected from numerical simulations, but have so far been rarely observed. Our systematic search for such galaxies in \textit{HST} images of the cores ($r{\lesssim}$0.5 Mpc) of massive clusters at $z{=}$0.3--0.5 from the MACS sample  revealed spectacular examples of ``jellyfish" galaxies undergoing dramatic transformations as the result of a high-speed encounter with the dense intra-cluster gas. 

We find the brightest of these galaxies ($M_{\rm F606W}<-$21) to be located closer to the cluster cores than would be expected for rapid stripping close to the ram-pressure stripping radius ($\sim$1 Mpc) upon first infall; in addition, their observed (projected) trajectories suggest that passages at large impact parameter are common. Many of these findings could be the result of selection effects. By design, our sample of the most spectacular cases of ram-pressure stripping is biased in favor of massive galaxies moving close to the plane of the sky.  Additional selection biases are plausible, acting against galaxies on radial orbits or infall paths, for which ram-pressure stripping that is as effective as predicted by theory and simulations would proceed too fast to be observed near the cluster core. If so, our search may have been efficient at selecting stripping events of particularly long duration in galaxies entering the cluster environment at grazing incidence, but penetrating the ICM to within the ram-pressure stripping radius. Detailed, spatially resolved investigations of current and past star formation as well as of the gas and stellar mass will be critical to elucidate the evolutionary history of these objects.

With very few systems (among them 235144$-$260358 in A\,2667) previously known in which ram-pressure stripping boosts the intrinsic luminosity of a galaxy to $M_{\rm F606W}{<}-$21, our sample of extreme ``jellyfish" galaxies in MACS clusters significantly increases the number of targets for studies of the physics and dynamics of the most  violent gas-galaxy interactions. At apparent magnitudes of $m_{\rm F606W}{\sim}$19, and featuring angular extents of typically $5-10^{\prime\prime}$, these galaxies also represent ideal targets for in-depth two-dimensional study with integral-field unit spectrographs \citep[e.g.,][]{2013MNRAS.429.1747M}. A more thorough investigation of the statistical properties of these systems (including their location and direction of motion within the host cluster) is underway, based on a greatly extended sample that includes much fainter jellyfish candidates.\\ \\

\begin{acknowledgments}
H.E.\ and L.N.S.\ gratefully acknowledge financial support from STScI grants GO-10491, -10875, \mbox{-12166}, and -12884. A.C.E.\ acknowledges support from STFC grant ST/I001573/1. We thank an anonymous referee for comments and suggestions that improved the comparison of our findings with those from previous work at lower redshift.
\end{acknowledgments}

\clearpage


\begin{thebibliography}{}

\bibitem[Abadi et al.(1999)]{1999MNRAS.308..947A} Abadi, M.~G., Moore, B., 
\& Bower, R.~G.\ 1999, \mnras, 308, 947 
\bibitem[Bell et al.(2004)]{2004ApJ...608..752B} Bell, E.~F., Wolf, C., 
Meisenheimer, K., et al.\ 2004, \apj, 608, 752 
\bibitem[Bertin 
\& Arnouts(1996)]{1996A&AS..117..393B} Bertin, E., \& Arnouts, S.\ 1996, \aaps, 117, 393 
\bibitem[Boselli et al.(2006)]{2006ApJ...651..811B} Boselli, A., Boissier, 
S., Cortese, L., et al.\ 2006, \apj, 651, 811 
\bibitem[Boselli \& Gavazzi(2006)]{2006PASP..118..517B} Boselli, A., \& Gavazzi, G.\ 2006, \pasp, 118, 517 
\bibitem[Boselli et 
al.(1997)]{1997A&A...327..522B} Boselli, A., Gavazzi, G., Lequeux, J., et al.\ 1997, \aap, 327, 522 
\bibitem[Byrd \& Valtonen(1990)]{1990ApJ...350...89B} Byrd, G., \& Valtonen, M.\ 1990, \apj, 350, 89 
\bibitem[Cayatte et al.(1990)]{1990AJ....100..604C} Cayatte, V., van 
Gorkom, J.~H., Balkowski, C., \& Kotanyi, C.\ 1990, \aj, 100, 604 
\bibitem[Cayatte et al.(1994)]{1994AJ....107.1003C} Cayatte, V., Kotanyi, 
C., Balkowski, C., \& van Gorkom, J.~H.\ 1994, \aj, 107, 1003 
\bibitem[Cortese et al.(2007)]{2007MNRAS.376..157C} Cortese, L., Marcillac, 
D., Richard, J., et al.\ 2007, \mnras, 376, 157 
\bibitem[Dressler(1980)]{1980ApJS...42..565D} Dressler, A.\ 1980, \apjs, 
42, 565 
\bibitem[Dressler 
\& Gunn(1983)]{1983ApJ...270....7D} Dressler, A., \& Gunn, J.~E.\ 1983, \apj, 270, 7 
\bibitem[Ebeling et al.(2001)]{2001ApJ...553..668E} Ebeling, H., Edge, 
A.~C., \& Henry, J.~P.\ 2001, \apj, 553, 668 
\bibitem[Faber et al.(2007)]{2007ApJ...665..265F} Faber, S.~M., Willmer, 
C.~N.~A., Wolf, C., et al.\ 2007, \apj, 665, 265 
\bibitem[Gavazzi(1989)]{1989ApJ...346...59G} Gavazzi, G.\ 1989, \apj, 346, 59 
\bibitem[Giovanelli \& Haynes(1985)]{1985ApJ...292..404G} Giovanelli, R., \& Haynes, M.P.\ 1985, \apj, 292, 404
\bibitem[Gunn \& Gott(1972)]{1972ApJ...176....1G} Gunn, J.~E., \& Gott, J.~R., III 1972, \apj, 176, 1 
\bibitem[Hester et al.(2010)]{2010ApJ...716L..14H} Hester, J.~A., Seibert, 
M., Neill, J.~D., et al.\ 2010, \apjl, 716, L14 
\bibitem[Kapferer et al.(2009)]{2009A&A...499...87K} Kapferer, W., Sluka, C., Schindler, S., Ferrari, C., \& Ziegler, B.\ 2009, \aap, 499, 87 
\bibitem[Le F{\`e}vre et al.(2000)]{2000MNRAS.311..565L} Le F{\`e}vre, O., 
Abraham, R., Lilly, S.~J., et al.\ 2000, \mnras, 311, 565 
\bibitem[Ma et al.(2008)]{2008ApJ...684..160M} Ma, C.-J., Ebeling, H., 
Donovan, D., \& Barrett, E.\ 2008, \apj, 684, 160 
\bibitem[Merluzzi et al.(2013)]{2013MNRAS.429.1747M} Merluzzi, P., 
Busarello, G., Dopita, M.~A., et al.\ 2013, \mnras, 429, 1747 
\bibitem[Moore et al.(1996)]{1996Natur.379..613M} Moore, B., Katz, N., 
Lake, G., Dressler, A., \& Oemler, A.\ 1996, \nat, 379, 613 
\bibitem[Moore et al.(1998)]{1998ApJ...495..139M} Moore, B., Lake, G., 
\& Katz, N.\ 1998, \apj, 495, 139 
\bibitem[Owen et al.(2006)]{2006AJ....131.1974O} Owen, F.~N., Keel, W.~C., 
Wang, Q.~D., Ledlow, M.~J., \& Morrison, G.~E.\ 2006, \aj, 131, 1974 
\bibitem[Owers et al.(2012)]{2012ApJ...750L..23O} Owers, M.~S., Couch, 
W.~J., Nulsen, P.~E.~J., \& Randall, S.~W.\ 2012, \apjl, 750, L23 
\bibitem[Scott et al.(2010)]{2010MNRAS.403.1175S} Scott, T.~C., 
Bravo-Alfaro, H., Brinks, E., et al.\ 2010, \mnras, 403, 1175 
\bibitem[Smith et al.(2010)]{2010MNRAS.408.1417S} Smith, R.~J., et al.\ 2010, \mnras, 408, 1417S
\bibitem[Steinhauser et al.(2012)]{2012A&A...544A..54S} Steinhauser, D., Haider, M., Kapferer, W., \& Schindler, S.\ 2012, \aap, 544, A54 
\bibitem[Sun et al.(2007)]{2007ApJ...671..190S} Sun, M., Donahue, M., \& Voit, G.~M.\ 2007, \apj, 671, 190 
\bibitem[Takeda et al.(1984)]{1984MNRAS.208..261T} Takeda, H., Nulsen, 
P.~E.~J., \& Fabian, A.~C.\ 1984, \mnras, 208, 261 
\bibitem[Toomre \& Toomre(1972)]{1972ApJ...178..623T} Toomre, A., \& Toomre, J.\ 1972, \apj, 178, 623 
\bibitem[Vollmer et al.(2000)]{2000A&A...364..532V} Vollmer, B., Marcelin, M., Amram, P., et al.\ 2000, \aap, 364, 532 
\bibitem[Vollmer et al.(2001a)]{2001A&A...374..824V} Vollmer, B., Braine, J., Balkowski, C., Cayatte, V., \& Duschl, W.~J.\ 2001a, \aap, 374, 824 
\bibitem[Vollmer et al.(2001b)]{2001ApJ...561..708V} Vollmer, B., Cayatte, V., Balkowski, C., \& Duschl, W.~J.\ 2001b, \apj, 561, 708 
\bibitem[Yagi et al.(2010)]{2010AJ....140.1814Y} Yagi, M., Yoshida, M., 
Komiyama, Y., et al.\ 2010, \aj, 140, 1814 
\bibitem[Yoshida et al.(2008)]{2008ApJ...688..918Y} Yoshida, M., Yagi, M., Komiyama, Y., et al.\ 2008, \apj, 688, 918 

\end{thebibliography}
\end{document}